\documentclass[12pt]{article}

\usepackage{graphicx}
\DeclareGraphicsExtensions{.pdf,.eps}

\begin{document}

%%% BEGIN ARTICLE

\title{Algorithmic Cooling of Spins: 
A Practicable Method for Increasing Polarization}

\author{Jos\'e M. Fernandez$^{1}$, Seth Lloyd$^{2}$, 
Tal Mor$^{3}$, and Vwani Roychowdhury$^{4}$
\\ \small 1. 
D\'epartement de g\'enie informatique, \'Ecole Polytechnique de Montr\'eal, Montr\'eal, Canada
\\ \small 2. 
Department of Mechanical Engineering, MIT, Cambridge,
USA
\\ \small 3. 
Computer Science Department, Technion, Haifa, Israel.
\\ \small 4. 
Electrical Engineering Department,  UCLA,
Los Angeles, CA, USA.
}

\date{21 January 2004}  

\maketitle

{\bf 
An efficient technique to generate ensembles of spins that 
are highly polarized by external magnetic fields is the 
Holy Grail in Nuclear Magnetic Resonance (NMR) spectroscopy. 
Since spin-half nuclei have steady-state polarization biases 
that increase inversely with temperature, spins exhibiting 
high polarization biases are considered {\em cool}, even 
when their environment is warm. Existing spin-cooling 
techniques~\cite{PT,Sorensen,SV99,Griffin} are highly 
limited in their efficiency and usefulness.
{\em Algorithmic cooling}~\cite{AC-PNAS} is a promising 
new spin-cooling approach that employs data compression 
methods in {\em open systems}. It reduces the entropy of  
spins on long molecules to a point far beyond Shannon's 
bound on reversible entropy manipulations (an information-theoretic
version of the 2nd Law of Thermodynamics),
thus increasing their polarization.
Here we present an efficient and experimentally feasible 
algorithmic cooling technique that cools spins to very 
low temperatures even on {\em short molecules}.
This {\em practicable algorithmic cooling} 
could lead to breakthroughs in high-sensitivity NMR 
spectroscopy in the near future, and to 
the development of scalable NMR quantum computers in the far future.
Moreover, while the cooling algorithm itself is classical, it uses 
quantum gates in its implementation, thus representing 
the first short-term application of quantum computing devices. 
}

\newpage

NMR is a technique used for studying
nuclear spins in magnetic fields~\cite{book-on-NMR}. 
NMR techniques 
are extremely useful in identifying 
and characterizing chemical
materials, potentially including materials that appear at 
negligible levels.
There are numerous NMR applications in biology, medicine, chemistry,
and physics, for instance magnetic resonance imaging 
(used for identifying malfunctions of various body
organs, for monitoring brain activities, etc.), identifying materials
(such as explosives or narcotics) for security
purposes, monitoring the purity of materials, and much more.
A major challenge in the application of NMR techniques 
is to overcome difficulties related to the 
Signal-to-Noise Ratio (SNR). 
Several potential methods have been proposed for improving the SNR,
but each of them has its problems and limitations. 
We thoroughly review in the Supplementary 
Information A the six main 
existing solutions to the SNR problem. 
In brief, the first (and not very effective)
three methods involve cooling the entire system,
increasing the magnetic field, and increasing the sample size.
The fourth method involves repeated sampling over time, 
a very feasible and commonly practised 
solution to the SNR problem. Its limitation is that 
in order to improve the SNR
by a factor of $M$, spectroscopy requires $M^2$ repetitions, 
making it an overly costly and
time-consuming method.

The fifth and sixth methods for improving the SNR
provide ways for cooling the spins 
without cooling the environment, which in this case refers to 
the molecules' degrees of freedom. 
This approach, known as ``effective cooling'' of the 
spins~\cite{PT, Sorensen, SV99}, 
is at the core of this current work, and is explained in more detail
in Supplementary Information A.  
The spins cooled by the effective cooling can be used 
for spectroscopy as long as they have
not relaxed back to their thermal equilibrium state.
For two-level systems 
the connection between the
temperature, the entropy, and the population probability
is a simple one. 
The population difference between these two levels
is known as the
polarization bias.
Consider a single spin-half particle
in a constant magnetic field. At equilibrium with a thermal heat bath
the probability of this spin to be up or down 
(i.e. parallel or anti-parallel
to the field's direction) is given by:
$P_{\uparrow}=\frac{1+\epsilon}{2}$,
and $P_{\downarrow}=\frac{1-\epsilon}{2}$.
The polarization bias,
$\epsilon=P_{\uparrow}-P_{\downarrow}$, is around 
$10^{-5}$---$10^{-6}$ in conventional NMR systems, so all
the following calculations are done to leading order in $\epsilon$.
A spin temperature at equilibrium is  
$T=\frac{\rm Const}{\epsilon}$.
A single spin can be viewed as a single bit with ``0" meaning spin-up
and ``1" meaning spin-down, and thus, the single spin 
(Shannon's) entropy is 
$H=1-(\epsilon^2/\ln 4)$.
A spin temperature out of thermal equilibrium 
is still defined via the same formulas.
Therefore, when a system is removed from thermal equilibrium, 
increasing the spins' polarization bias is equivalent to
cooling the spins (without cooling the system)
and to decreasing their entropy. 

One of these latter two
methods for effective cooling of the spins is the 
{\em reversible polarization compression (RPC)}. It is based on 
entropy manipulation techniques (very similar to data
compression), 
and can be used to cool some spins (bits) while heating 
others~\cite{Sorensen,SV99}.
Contrary to conventional data compression, 
RPC techniques focus on the 
low-entropy bits (spins),
namely those that get colder during the entropy manipulation
process. 
The other effective cooling method is known as
{\em polarization transfer}~\cite{PT}:
If at thermal equilibrium at 
a given temperature, the spins we
want to use for spectroscopy (namely, the observed spins) 
are less polarized than other spins (namely, auxiliary spins),
then transfering polarization from the auxiliary 
highly polarized spins into the observed spins  
is equivalent to cooling the observed spins, while heating
the auxiliary spins. 
In its most general form the RPC can be applied to spins 
which have different initial polarization biases. It follows 
that polarization transfer is merely a special case of RPC.  
Therefore, we refer to these two techniques together as RPC.

\begin{figure}[ht] 
\begin{center}
\includegraphics[scale=0.5]{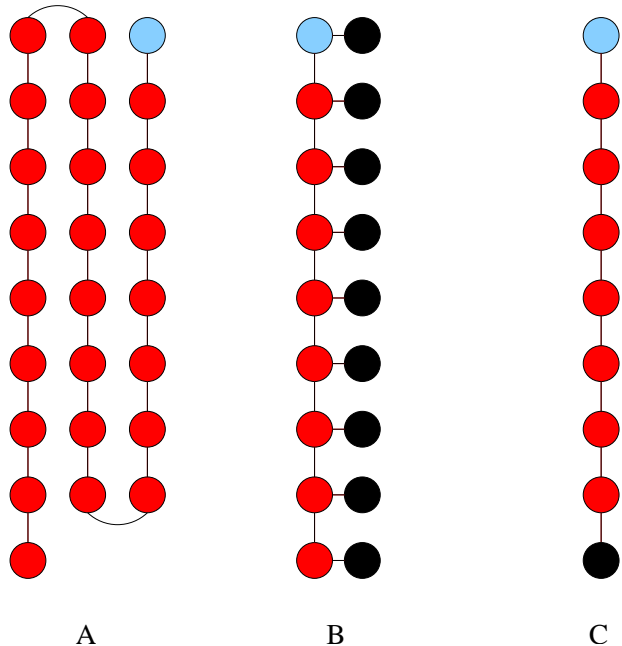}\end{center}
\caption{{\em Comparing space complexity of 
RPC and practicable 
algorithmic cooling.}
We present here the number of spins required in order to
improve the polarization bias of a single spin 5 times 
via RPC and via 
practicable algorithmic cooling 1 and 2,
when all spins have the same 
initial polarization bias
$\epsilon$. 
The colors illustrate the temperature, 
black being the thermal equilibrium temperature, 
blue being colder (five times colder, in the case we present), 
and red being hotter. 
Case A (RPC):  
the number of required spins is calculated from 
Shannon's bound~(Eq.\ref{EQ:Shannon-bound}) --- 
25 spins are required.
[Note that any specific algorithm shall 
probably be much worse than this theoretical
bound, requiring many more spins.]
Case B (PAC1):
the algorithm ``PAC1'' (for Practicable
Algorithmic Cooling 1) is the algorithm described in the text, 
and summarized 
in~Eq.~\ref{EQ:PAC1-Algo} ---
18 spins are required.
Note that with one additional single (parallel) PT step and one 
additional RESET step all hot spins regain their initial bias.
Case C (PAC2):
the algorithm ``PAC2'' (for Practicable
Algorithmic Cooling 2) is described in 
Supplementary Information E ---
9 spins are required.
[A remark.
In order to improve the polarization 25 times the numbers are:
RPC --- 625 spins are required;
PAC1 --- 34 spins are required;
PAC2 --- 17 spins are required. We did not provide figures due to
the difficulty of providing the figure for the RPC case.]
}
\label{Fig:compare-methods}
\end{figure}

RPC can be understood as reversible, in-place, lossless, adiabatic
entropy manipulations in a closed system. 
Unfortunately, as in data compression, 
RPC techniques are limited in their cooling
applicability because Shannon's bound states that entropy 
in the closed system cannot decrease.  
[Shannon's bound on entropy manipulations is an 
information-theoretic version of the 2nd Law of Thermodynamics.]
Polarization transfer is limited because
the increase of the polarization bias 
is bounded by the spin polarization bias of the
auxiliary highly-polarized spins. 
RPC done on $n$ uncorrelated spins with equal biases requires  
extremely long 
molecules in order to provide significant cooling.
The total
entropy of such spins satisfies
$H(n) = n(1-\epsilon^2/\ln 4)$.  Therefore, 
with $\epsilon=10^{-5}$ for instance, molecules
of length of an order of magnitude of $10^{10}$ 
are required in order to cool 
a single spin close to zero temperature.
For more modest (but still significant)
cooling, one can use smaller molecules 
(with $n \ll 10^{10}$) and 
compress the entropy into $n-1$ fully-random spins. 
The entropy of the remaining single spin satisfies
$H({\rm single})\ge 1-n\epsilon^2/\ln4$,
thus we can at most improve its polarization to
\begin{equation}\label{EQ:Shannon-bound}
\epsilon_{\rm final}\le\epsilon\sqrt{n}\ .
\end{equation}
Figure~\ref{Fig:compare-methods}A provides an illustration
of how Shannon's bound limits RPC techniques.
Unfortunately, manipulating many spins, say  $n>100$,
is a very difficult task, and the gain of $\sqrt{n}$
in polarization here is not 
nearly substantial enough to justify putting this technique  
into practice. 

We and our colleagues (Boykin, Mor, Roychowdhury, Vatan and 
Vrijen, hereinafter refered to as BMRVV), invented 
a promising new spin-cooling technique which we call 
{\em Algorithmic cooling}~\cite{AC-PNAS}.
Algorithmic cooling expands the effective cooling techniques 
much further by exploiting entropy manipulations in {\em open systems}. 
It combines RPC with relaxation (namely, thermalization)
of the {\em hotter spins},
as a way of pumping entropy outside the system and cooling the system
{\em much beyond Shannon's bound}. 
In order to pump the entropy out of the system,
algorithmic cooling employs regular spins (which we
call computation spins) together with rapidly relaxing spins. 
Rapidly relaxing spins are auxiliary spins that return to their 
thermal equilibrium state very rapidly. We refer to these 
spins as reset spins, or, equivalently, reset bits. 
The ratio, $R_{\rm relax-times}$, 
between the relaxation time of the 
computation spins and the relaxation time of the reset spins must
satisfy $R_{\rm relax-times} \gg 1$. 
This is vital if we wish to perform many cooling
steps on the system. 

The BMRVV algorithmic cooling  
used blocks of $m$ computation bits and pushed 
cooled spins to one
end of such blocks in the molecule. 
To obtain significant cooling, the algorithm required very long molecules
of hundreds or even thousands of spins, 
because its analysis relied on the law
of large numbers. 
As a result, although much better than any RPC technique, 
the BMRVV algorithmic cooling was still far from having  
any practical implications.

In order to overcome the need for large molecules, 
and at the same time capitalize on the great advantages 
offered by algorithmic cooling, we searched for new types 
of algorithms that would not necessitate the use of the 
law of large numbers for their analysis.  
Here we present examples of a novel, efficient, 
and experimentally feasible algorithmic cooling 
technique which we name ``{\em practicable algorithmic cooling}''. 
The space requirements of practicable algorithmic cooling are much 
improved relative to the RPC, 
as we illustrate in Figures~\ref{Fig:compare-methods}B 
and~\ref{Fig:compare-methods}C.
In contrast to the BMRVV algorithm,  
practicable algorithmic cooling techniques  
determine 
in advance the polarization bias of each cooled spin at each step of 
the algorithm. We therefore do not need to make use of the 
law of large numbers in the analysis of these techniques,
thus bypassing the shortcomings of 
the BMRVV algorithmic cooling.  Practicable algorithmic cooling 
{\em allows the use of very short molecules} to achieve the 
same level of cooling as the BMRVV achieves using much 
longer molecules, e.g., 34 spins instead of 180 spins, 
as we demonstrate explicitly in 
Supplementary Information B.

Both RPC and algorithmic cooling 
can be understood as applying a set of logical
gates, such as NOT, SWAP etc., onto the bits. 
For instance, one simple way to obtain polarization transfer
is by a SWAP gate. 
In reality, spins correspond to quantum bits.
Here we provide 
a simplified ``classical'' explanation, and the 
more complete ``quantum'' explanation is provided 
in Supplementary Information C.
A molecule with
$n$ spins can represent an $n$-bit 
computing device.  
A macroscopic number of identical molecules is available in a bulk
system, and these molecules act as though they are 
many computers performing the
same computation in parallel. 
To perform a desired computation, the same sequence
of external pulses is applied to all the molecules/computers. 

Algorithmic cooling 
is based on combining three very different operations: 
\begin{enumerate}
\item \label{enu:compress} RPC steps 
change the (local) entropy in the system so that some computation bits
are cooled while other computation bits become much hotter 
than the environment.
\item \label{enu:swap} Controlled interactions allow the hotter 
computation bits to
adiabatically lose their entropy to a set of reset bits via
polarization transfer from the reset bits into these specific
computation bits. 
\item \label{enu:wait} The reset bits 
rapidly return to their initial
conditions and convey their entropy to the environment, while the
colder parts (the computation bits)
remain isolated, so that the entire system is cooled.
\end{enumerate}
By repeatedly alternating between these operations, 
and by applying
a recursive algorithm, spin 
systems can be cooled to very low temperatures.

\begin{figure}[ht] 
\begin{center}
\includegraphics[scale=0.7]{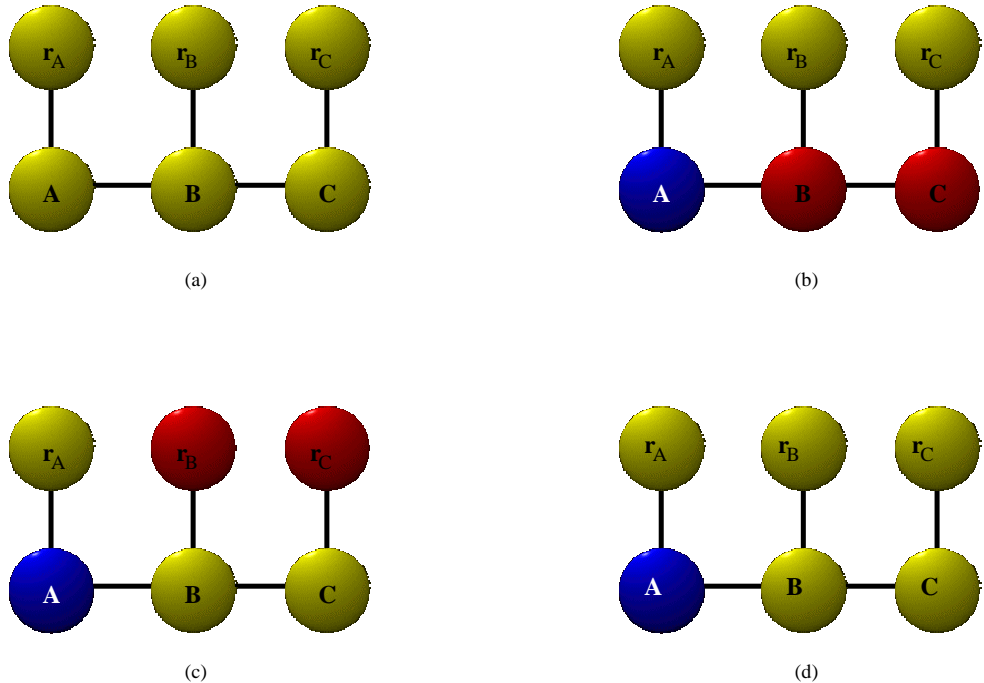}\end{center}
\caption{{\em A molecule for running a simplified algorithm.}
We show here an abstract example of a molecule with three computation 
bits $ABC$,
such that each is attached to a reset bit. All these bits have
the same polarization bias. 
The temperature after each step is illustrated by the colors:
yellow means the temperature at thermal equilibrium;
blue means colder than the initial temperature; 
and red means hotter than the initial temperature.
Let three computation bits $A$, $B$, and $C$ be given, 
such that each is attached
to a reset bit ($r_A$, $r_B$, and $r_C$), and {\em all bits}
have the same bias $\epsilon_{0}$ (see
Figure~\ref{Fig:ABCmolecule}a).
Our algorithm uses a particular 
RPC step on 3 bits
that we call 3-bit compression (3B-Comp), 
Polarization Transfer (PT) steps, and RESET steps:
1.--- 3B-Comp($A;B;C$);
the outcome of this step is shown in 
Figure~\ref{Fig:ABCmolecule}b.
2.---
PT($r_B \rightarrow B$), PT($r_C \rightarrow C$);
the outcome of this step is shown in 
Figure~\ref{Fig:ABCmolecule}c.
3.---
RESET($r_B,r_C$);
the outcome of this step is shown in 
Figure~\ref{Fig:ABCmolecule}d.
The 3-bit-compression used in the first step operates on the 
three computation bits, increasing the bias of bit $A$ by $3/2$
times, while heating the other two spins. Note that in the last 
step $r_A$ also relaxes, but is not modified by the RESET operation.
As a result of the 3B-Comp step that cools bit $A$, 
and the later PT and RESET steps
that cause all the other bits to regain their initial biases,
the entire system is cooled down. 
}
\label{Fig:ABCmolecule}
\end{figure}

In our simplest practicable algorithm 
there is an array of 
$n$ computation bits, and
each computation bit $X$ has a neighbouring reset bit $r_X$ 
with which it can be swapped. The algorithm uses 
polarization transfer steps, reset steps and 
the 3-bit RPC step which we call
3-bit-compression (3B-Comp). 
The 3B-Comp   
can be built, for instance, via the following two gates:
1.---Use bit $C$ as a control, and bit $B$ as a target; apply a CNOT
(Controlled-NOT) operation: 
$C\rightarrow C$, $B\rightarrow B\oplus C$,
where $\oplus $ denotes a logical eXclusive OR (XOR).
2.---Use bit B as a control, 
and bits A and C as targets; apply a variant
of a C-SWAP operation: $A\rightarrow C\bar{B}+AB$, $B\rightarrow B$,
$C\rightarrow A\bar{B}+CB$; 
this means that $A$ and $C$ are swapped if $B=0$. 
The effect of a 3B-Comp step in case it is applied
onto three bits with equal bias $\epsilon \ll 1$ is to
increase the bias of bit $A$ to
\begin{equation}\label{improved-bias} 
\epsilon_{\rm new} = \frac{3 \epsilon}{2} \ .
\end{equation}
See Supplementary Information D for calculation details.
A simplified example for practicable algorithmic cooling
based on these three steps and using $n=3$ computation bits
is presented in 
Figure~\ref{Fig:ABCmolecule}.

\begin{figure}[ht] 
\begin{center}
\includegraphics[scale=0.4]{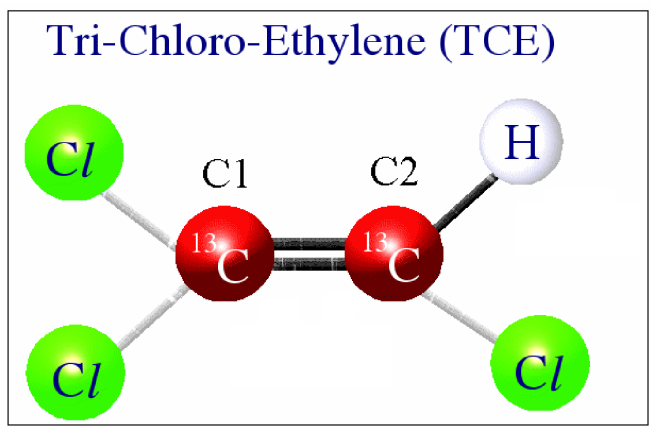}\end{center}
\caption{{\em  
A 3-bit computer:
a TCE molecule labeled with two $^{13}$C.}
The TCE molecule 
has three spin-half nuclei, the two Carbons and the Hydrogen,
which are given the names 
C1, C2 and H, as portrayed in the figure. 
Only these spins take a non-negligible part in the process, 
so the TCE molecule can therefore act as a three-bit computer.
The Chlorines have a very
small signal and their coupling with the Carbons can be averaged out
so they
cannot be considered as bits.
TCE can be used for polarization transfer 
because the Hydrogen's nucleus is four times
more polarized than the Carbons' nuclei.
The Hydrogen can also be used as a reset bit, because its relaxation
time can be made much shorter than the relaxation time of the Carbons. 
Based on the theoretical ideas presented in this current work,
we and colleagues~\cite{Cool-by-therm,YW} 
used the Hydrogen in the TCE to cool both spins C1 and C2,
decrease the total entropy of the molecule,
and bypass Shannon's bound on cooling
via RPC.
}
\label{Fig:TCE}
\end{figure}

In many realistic cases 
the polarization bias of
the reset bits at thermal equilibrium
is higher than the polarization bias
of the computation bits. 
By performing 
an initiation process of polarization
transfer from relevant reset bits to relevant 
computation bits,
prior to any polarization compression, we cool the computation bits
to the 
initial bias of the reset bits,
$\epsilon_0$, the zero'th purification level.
See for instance 
Figure~\ref{Fig:TCE}.

In the algorithm we now present,
we attempt to cool a single
bit much more rapidly
than possible in the reversible processes.
In order to cool a single bit (say, bit $A$) 
to the first purification level,
$\epsilon_1=3\epsilon_0 / 2$, start with three computation 
bits, $ABC$, perform a Polarization Transfer (PT) 
step to initiate them, 
and perform 3B-Comp to increase the
polarization of bit $A$.
In order to cool one bit (say, bit $A$) to the second purification
level (polarization bias $\epsilon_{2}$), 
start with 5 computation bits $(ABCDE)$. 
Perform a PT step in parallel to initiate bits $ABC$, 
followed by a 3B-Comp that
cools $A$ to a bias $\epsilon_{1}$. 
Then repeat the above (PT + 3B-Comp)
on bits $BCD$ to cool bit $B$, 
and then on bits $CDE$ to cool bit $C$ as well. 
Finally, apply 3B-Comp to bits $ABC$ to purify bit
$A$ to the second purification level 
$\epsilon_2$,
which for small biases gives $\epsilon_2 \approx (3/2)\epsilon_1
\approx (3/2)^2 \epsilon_0$. 

This (very simple) practicable algorithmic cooling,
which uses the 3B-Comp,  PT, and RESET steps,
is now easily generalized to cool a single 
spin to any purification level $J_f$. 
Let $j$ be an index telling the polarization level
$j \in {1 \ldots J_f}$.
To obtain one bit at a purification level $j$ we define the procedure
$M_{j}$. 
Let $M_{0}(A)$ be the PT step
from a reset bit $r_A$ to bit $A$ to yield a 
polarization bias $\epsilon_{0}$. Then the procedure
$M_{1}$ contains three such PT steps
(performed in parallel on three neighbouring
bits) followed by one 3B-Comp that cools the left bit (among the three)
to level 1. 
In order to keep track of the locations of the cooled
bit we mention here the location of the cooled bit. 
For an array of $n$
bits, $a_{n}a_{n-1}\ldots a_{2}a_{1}$ we write $M_{1}(k)$ to say
that $M_{1}$ is applied on the three bits $a_{k};a_{k-1};a_{k-2}$
so that bit $a_{k}$ is cooled to a bias level 1. 
Similarly, $M_{2}(k)$ means that $M_{2}$ is applied on the five bits 
$a_{k};a_{k-1}; \ldots;a_{k-4}$, 
so that bit $a_{k}$ is cooled to a bias level $2$. 
We use the notation ${\mathcal{B}}_{\{(j-1)\rightarrow j\}}(k)$
to present the 3B-Comp applied onto bits 
$a_{k};a_{k-1};a_{k-2}$, 
purifying bit $a_{k}$ from $\epsilon_{j-1}$
to $\epsilon_{j}$. Then, the full algorithm has a simple recursive
form: for $j\in \{1,\ldots ,J_{f}\}$
\begin{equation}
M_{j}(k)={\mathcal{B}}_{\{(j-1)\longrightarrow j\}}(k)
M_{j-1}(k-2)\; M_{j-1}(k-1)\; M_{j-1}(k) \ ,
\end{equation}
applied from right
to left ($M_{j-1}(k)$ is applied first). 
For instance, 
$M_{1}(3)={\mathcal{B}}_{\{0\longrightarrow 1\}}(3)
M_{0}(1)\; M_{0}(2)\; M_{0}(3)$
is the 3B-Comp applied after reset. 
The procedure cooling one bit to
the second level (starting with five bits) is written as 
$M_{2}(5)={\mathcal{B}}_{\{1\longrightarrow 2\}}(5)
M_{1}(3)\; M_{1}(4)\; M_{1}(5)$.
Clearly, $M_{1}(k)$ can be applied 
to any $k \ge 3$, $M_{2}(k)$ can be applied to any 
$k \ge 5$, and $M_{j}(k)$ can be applied to any 
$k \ge 2j+1$.
Thus, if we wish to bring a single bit to a polarization level $J_{f}$, 
then $2J_{f}+1$ computation bits and $2J_{f}+1$ reset
bits are required. 
We conclude that the algorithm uses 
\begin{equation} \label{EQ:PAC1-Algo}
4J_f + 2
\end{equation}
bits.
Due to Eq.(~\ref{improved-bias}) the final polarization bias is
$\epsilon_{\rm final} = \epsilon_{J_f} 
\approx (3/2)^{J_f}\epsilon_0$,
as long as $\epsilon_{J_f} \ll 1$. 

Let us now calculate the time complexity of the algorithm.
The number of time steps
in operation $M_j$ presented above 
is $T_j$ with $T_0 = 1$, $T_1 = 2$ (since the three
resets are done in parallel), 
$T_2 = 7$, and $T_{j} = 3 T_{j-1} + 1$,
for any $j \ge 2$.
The recursive formula yields 
$T_{j} = 3^{j-1} T_1 + (3^{j-1}-1)/2$,
so that finally
\begin{equation}
T_{J_f} = \frac{5 \times 3^{J_f - 1} -1}{2}
\ . \label{timesteps}
\end{equation}
If we count the reset steps only, assuming a much 
longer time for reset steps than for the other steps, then we get
$T_{J_f} ({\rm reset}) = 3^{J_f -1}$ reset time steps.
In Supplementary Information B  
we compare the space and time complexity
of this practicable algorithmic cooling 
with the BMRVV algorithmic cooling.

If we use the reset bits
also for the computation, we can improve the space complexity by 
a factor of 2. The algorithm then uses
$2J_{f}$ computation bits and one reset bit, thus a total of 
$2J_f + 1$ bits.
See Supplementary Information E 
for a full description of this algorithm.
In Figure~\ref{Fig:compare-methods} we compare the two algorithms 
descibed here with the RPC, 
to explicitly illustrate the advantages of practicable algorithmic 
cooling.

We suggested here an algorithmic 
cooling technique which is feasible with existing
technology. 
A variant of the 3B-Comp step was implemented by~\cite{BCS-Exp} 
following Schulman and Vazirani's
RPC (their molecular-scale heat engine)~\cite{SV99}.
The PT plus RESET steps have been implemented 
by us and our colleagues~\cite{Cool-by-therm,YW} 
following the theoretical ideas presented in this
current work. 
Thus, all steps needed for implementing practicable algorithmic cooling 
of spins have 
already been successfully 
implemented in NMR laboratories. 
Both these implementations use Carbon molecules
that contain two enriched $^{13}C$ atoms 
which have spin-half nuclei. 
In addition to the Carbons, the 3B-Comp experiment~\cite{BCS-Exp} 
also uses one Bromine nucleus,
the 3-bit molecule
being $\mathrm{C}_{2}\mathrm{F}_{3}\mathrm{Br}$. The  
compression yields a single cooled bit with a polarization bias 
increased by a factor of 1.25. 
The experimental PT and RESET steps~\cite{Cool-by-therm,YW}
use one Hydrogen and the two Carbons, and the 3-bit molecule 
is the 
Trichloroethylene (TCE) molecule (see Figure~\ref{Fig:TCE}),
with the two Carbons being the computation bits 
and the Hydrogen being the reset bit. 
This work of~\cite{Cool-by-therm,YW} implements our ideas 
in order to prove experimentally that it is possible to
bypass Shannon's
bound on entropy manipulations. 
It is important to note that cooling spins to high purification 
levels is a feasible task but certainly not an easy one. 
Addressing specific
spins in molecules containing many spins is a very challenging 
operation
from an experimental point of view. 
Another important challenge is achieving a sufficiently good 
ratio $R_{\rm relax-times}$ that will allow the performance 
of many reset operations, before the cooled bits naturally
re-thermalize.

As demonstrated here the theoretical practicable algorithmic cooling
is a purely classical algorithm. However, the building blocks
are spin-half particles which are quantum bits, 
and NMR machine pulses which implement
quantum gates, that have no classical analogues.
Practicable algorithmic cooling is, therefore, the {\em first 
short-term application} of quantum computing 
devices${}^{(}$\footnote{Practicable
algorithmic cooling is also one of the first short-term 
applications of quantum
information processing}${}^{)}$, 
using simple quantum computing devices to improve SNR in 
NMR spectroscopy and imaging. Both practicable algorithmic cooling
and quantum key distribution could lead to implement

Algorithmic cooling could also have an important {\em long-term
application}, namely, quantum computing 
devices that can run 
important quantum algorithms such as factorizing large numbers~\cite{Shor}. 
NMR quantum computers~\cite{NMR-exp1,NMR-exp2} 
are currently the most successful quantum computing devices
(see for instance,~\cite{factoring-15}), 
but are known to suffer from severe 
scalability problems~\cite{scaling-problem1,scaling-problem2, AC-PNAS}. 
Impressive theoretical solutions were provided in~\cite{SV99,AC-PNAS} 
but only   
the algorithms presented here can lead to a realistic solution.
As explained in
Supplementary Information A1, practicable algorithmic cooling 
can be used for building scalable NMR quantum computers
of 20-50 quantum bits 
if electron spins will be used for the PT and RESET steps.

\clearpage

\par{\bf Acknowledgements:}

We would like to thank Ilana Frank for very thorough editing of this
paper, and to thank Gilles Brassard, Ilana Frank, and Yossi Weinstein
for many helpful comments and discussions.  T.M. thanks the Israeli
Ministry of Defense for supporting a large part of this research. His
research was also supported in part by the promotion of research at
the Technion, and by the Institute for Future Defense Research.  This
research was done while J.M.F. was at the Universit\'e de Montr\'eal.

\par{\bf Competing interests and materials request statements:}

Practicable algorithmic 
cooling has led to a patent application, US number 
60/389,208.

Correspondance and requests for material should be addressed to Jos\'e
M. Fernandez.  Email:jose.fernandez@polymtl.ca.

\clearpage

\clearpage

\clearpage

%%% END ARTICLE

%%%%%%%% BEGIN SUPPLEMENTARY MATERIAL

\newpage

%\section*{Supplementary On-line Material}
\section*{Supplementary Material}

\appendix

\section{Solutions to the SNR problem and their limitations}
\label{APP:SNR-solve}
There are several methods which provide rather trivial 
solutions to the SNR problem.
The first of these involves simply cooling
the entire system, including cooling the spins, thereby
improving the SNR.
This is not usually a useful solution 
because cooling the system
can result in modifying the
inspected molecules (e.g., solidifying the material), 
or destroying the sample 
(e.g., killing the patient).
A second method involves increasing the magnetic field, 
which in some sense
is equivalent to cooling the system. This solution is generally
much too expensive
to be considered feasible, 
as it adds tremendously to the cost of the necessary machinery.
The third method involves increasing the sample size, 
such as taking a larger blood sample.
It is usually
impossible, however, to implement this solution 
due to the large sample size 
that would be required in order to improve the SNR.
One would need to increase the sample size by a factor of
$M^2$ in order to improve the SNR
by a factor of $M$. In the case of a blood sample, then, 
to obtain a one hundredfold improvement in SNR one would 
need to take 100,000 ml. of blood instead of 10 ml., 
which is clearly not feasible.
In some cases increasing the sample size 
is strictly impossible, 
for example if the sample in question is a human brain.
NMR machine size limitations also severely limit the applicability
of this method.
The fourth method involves repeated sampling over time. 
This is a very feasible and commonly practised 
solution to the SNR problem. Its limitation is that 
in order to improve the SNR
by a factor of $M$, spectroscopy requires $M^2$ repetitions, 
making it an overly costly and
time consuming method.
In many cases it is also simply impractical, due to 
changes over time in the sample during the spectroscopy 
process (e.g., changes in concentration of
a material in a certain body organ). 
The long recovery time between measurements also contributes 
to the impracticality of this method. Furthermore, this 
solution is less useful if the noise in question 
is not a Gaussian noise.

An additional two methods provide ways for cooling the spins 
without cooling the environment, which in this case refers to 
the molecules' degrees of freedom. 
This approach is known as ``effective cooling'' of the spins.
Such cooling is as good as regular cooling of the system, because
the cooled spins can be used for spectroscopy as long as they have
not relaxed back to their thermal equilibrium state.
For k-level quantum systems there is, generally, 
a complex relationship between
temperature, entropy, and population density at the different levels.
For the sake of simplicity we shall focus here only on
two-level systems, namely spin-half particles. 
Note, however, 
that the results here can be generalized also to
higher level systems, and to quantum systems other than spins.
For two-level systems 
the connection between the
temperature, the entropy, and the population density 
is a simple one. 
In two-level systems the population density difference
is known as the
polarization bias.
Cooling the system and therefore also the spins (after 
they reach thermal equilibrium)
results in increasing the spins'
polarization bias,
and decreasing their entropy.
Consider a single spin
in a constant magnetic field. At equilibrium with a thermal heat bath
the probability of this spin to be up or down 
(i.e. parallel or anti-parallel
to the field's direction) is given by:
$P_{\uparrow }=\frac{1+\epsilon }{2}$,
and $P_{\downarrow }=\frac{1-\epsilon }{2}$.
The polarization bias,
$\epsilon = P_\uparrow - P_\downarrow $, is calculated to be 
$\epsilon =\tanh \left(\frac{\Delta E }{2K_{B}T}\right)$,
where $\Delta E$ is the energy gap between the up and down states
of the spin, $K_{B}$ is Boltzman's coefficient 
and $T$ is the temperature
of the thermal heat bath. 
Note that $\Delta E = 2 \gamma B$ with $B$ the magnetic field, 
and $\gamma$ the material dependant gyromagnetic constant which depends on the nucleus or particle, and is thus responsible
for causing the differences in polarization biases
[the ${}^{13}$Carbon's nucleus, for instance, is 4 times more 
polarized than Hydrogen's nucleus; 
electron-spin is about $10^3$ times more polarized than Hydrogen's nucleus.]
For high temperatures or small biases we approximate
\begin{equation} \label{polar-temp}
\epsilon \approx 
\frac{\Delta E}{2K_{B}T}
\end{equation} 
to leading order.
A spin temperature at equilibrium is therefore 
$T=\frac{\Delta E}{2K_{B} \tanh^{-1}(\epsilon)}$, which is
$T \approx \frac{\Delta E}{2\epsilon K_{B}}$
to leading order (for small $\epsilon$).
A single spin can be viewed as a single bit with ``0" meaning spin-up
and ``1" meaning spin-down. Then, the single spin 
entropy (for a small polarization bias) is 
\begin{equation} \label{EQ:entropy}
H= - P_\uparrow \log_2 P_\uparrow - P_\downarrow \log_2 P_\downarrow
\approx 1 - (\epsilon^2 / \ln 4) + O(\epsilon^4) \ .
\end{equation} 
A spin temperature out of thermal equilibrium 
is still defined via the same formulas.
Therefore, when a system is removed from thermal equilibrium, 
increasing the spins' polarization bias is equivalent to
cooling the spins (without cooling the system)
and to decreasing their entropy. 
What we have learned about single spins can be generalized
to several spins on a single molecule.

One method for effective cooling of the spins
is the 
{\em reversible polarization compression (RPC)}. 
As is explained in the paper, RPC
techniques essentially
involve harnessing the powerful data compression tools which 
have been developed over the past several decades 
and making use of them to cool spins, as much as 
is allowed by Shannon's bound on entropy manipulations
in a closed system~\cite{SV99}. 
Let us present in more details the 
example of how Shannon's bound limits RPC techniques.  The total
entropy of the $n$ uncorrelated spins with equal biases satisfies
$H(n) = n(1-\epsilon^2/\ln 4)$. 
This entropy could be compressed 
into $m \ge n(1-\epsilon^2)/\ln4$ 
high entropy spins, 
leaving $n - m $ extremely cold spins that have almost zero entropy.
Due to the preservation of entropy, 
the number of extremely cold bits
cannot exceed $n \epsilon^2/\ln4$.
With $\epsilon=10^{-5}$ for instance, extremely long molecules
whose length is of an order of magnitude of $10^{10}$ 
are required in order to cool 
a single spin close to zero temperature.
If we use smaller molecules, with $n \ll 10^{10}$, and we
compress the entropy into $m=n-1$ fully-random spins, 
the entropy of the remaining single spin satisfies
$H({\rm single\ spin}) \ge n(1 - \epsilon^2 / \ln4) - (n-1)
= 1 - n\epsilon^2/\ln4 $.
Thus we can at most reduce its entropy to  
$1 - [\epsilon^2_{\rm final} / \ln4] \ge 1 -  n\epsilon^2/\ln4$,
so that its polarization bias is improved to  
\begin{equation} 
\epsilon_{\rm final} \le \epsilon \sqrt n \ .
\end{equation}
Unfortunately, manipulating many spins, say  $n>100$,
is a very difficult task, and the gain of $\sqrt{n}$
in polarization here is not 
nearly substantial enough to justify putting this technique  
into practice. 
It is interesting to note that improving the SNR
by a factor of $M$ 
can either be performed using $n=M^2$ spins via RPC, 
or using $M^2$ repetitions
via repeated sampling
over time, or using $M^2$ samples at once, 
via increasing the sample size times $M^2$.

The other effective cooling method is the special case of RPC called
{\em Polarization transfer}.
Polarization transfer is limited as a cooling technique because
the increase of the polarization bias of the observed spins
is bounded by the spin polarization bias of the
highly polarized auxiliary spins. The polarization transfer 
method is commonly implemented by 
transfering the polarization 
among nuclear spins on the
same molecule. 
This form of implementation is regularly used in NMR
spectroscopy, and it provides, say,   
an increase of about one order of magnitude.
As a simple demostration, consider for example, 
the 3-bit molecule trichloroethylene
(TCE) shown in Figure~3 in the paper. 
The Hydrogen's nucleus is four times more polarized than 
each of the Carbons' nuclei,
and can be used via polarization transfer 
to cool a single Carbon 
(say, the Carbon reffered to as C2),
by a factor of four.

A different form of polarization transfer involves 
removing entropy from the nuclear spins into electron spins or 
into other molecules. 
This technique is still at initial stages of 
its developement~\cite{Griffin},
but might become extremely important in the future as is explained 
in the Discussion of the paper and in Appendix~\ref{APP:PT}.

\subsection{Polarization transfer with electron spins} \label{APP:PT}

Suppose that performing polarization transfer 
with spins other than the nuclear spins 
on the same molecule
(e.g., spins on other types of molecules, 
or electron spins on the same molecule)
were to become feasible. 
It would clearly open interesting options potentially 
leading to a much more impressive
effective cooling than the regular polarization transfer 
onto neighboring nuclear spins.  Polarization transfer
with electron spins~\cite{Griffin} 
could lead to three and maybe even four 
orders of magnitude
of polarization increase. Unfortunately, severe technical
difficulties with the implementation of this method, such as the 
need to master two very different 
electromagnetic frequencies within one machine, 
have so far prevented it from 
being adopted as a common practice. 
However, if such polarization transfer steps 
come into practice, and if the same machinery
were to allow conventional NMR techniques as well,
then algorithmic cooling could be applied with much better parameters.
First, $\epsilon_0$ could be increased to
around $0.01$ or even $0.1$. 
Second, the ratio $R_{\rm relax-times}$ could easily
reach $10^3$ or maybe even $10^4$.
With these numbers, scalable quantum computers of 20-50 bits 
might become feasible, as can be seen in the tables
presented in Appendix~\ref{APP:compare-AC1}.

\section{Comparison with the original algorithmic cooling} 
\label{APP:compare-AC1}

In order to compare our practicable algorithmic cooling algorithms 
with the BMRVV
algorithmic cooling we need to consider longer molecules, and 
more time steps than considered in the text. 
This clearly makes the schemes unfeasible with existing
technology. However, if such parameters 
become feasible, for instance using the
techniques described in Appendix~\ref{APP:PT}, then medium-scale
quantum computing devices of size 20-50 quantum bits can be built.
It is important to mention that a comparison of the BMRVV
algorithmic cooling with algorithims for 
RPC is provided in the supplementary
material (Appendix C) of~\cite{AC-PNAS},
to clearly show the advantage of
the BMRVV algorithmic cooling over RPC. 

In the following, we use the structure in which there are $n$
computation bits and a reset bit attached to each computation bit.
For any desired final number of cooled bits $m$, 
the algorithm described in
the text, namely PAC1 (Practicable Algorithmic Cooling 1), 
requires 
$n= 2 J_f + m$
computation bits, 
and the number of total time steps is $m T_{J_f}$.
These numbers are obtained as trivial generalizations of 
Eq.~4 and~5 in the paper. 
We compare the performance
of our new algorithmic cooling and the 
BMRVV algorithmic cooling~\cite{AC-PNAS}
in a case where the goal is to obtain 20 extremely cold bits. 

To obtain 20 cooled bits via 
Practicable Algorithmic Cooling 1 (PAC1)
we need to use 
$n = 2 J_f + 20$ computation bits and 
$T <  50 \times 3^{J_f - 1}$ 
time steps
This can be compared with~\cite{AC-PNAS} where 
$n \approx 40 j_f$ [Eq.(9), with $\ell=5$] 
and $T < 400 \times 5^{j_f+1}$ [Eq.(10), with $\ell=5$].
However, the required $j_f$ of~\cite{AC-PNAS} is different from 
the required $J_f$
of the current work, 
because $\epsilon_j \approx 2 \epsilon_{j-1}$ in~\cite{AC-PNAS}
while 
$\epsilon_j \approx (3/2) \epsilon_{j-1}$ here (see
Eq~2 in the paper).
As a result, a smaller $j_f$ is required 
in the algorithm of~\cite{AC-PNAS}, if a similar desired 
polarization bias is to be obtained.
Tables~\ref{Tab:AC1} and~\ref{Tab:AC2} present
a fair comparison between 
the time-space requirements of the original and the new algorithms.
As can be seen, the improvement obtained by our  
practicable algorithmic cooling (PAC1)
is very impressive, both in space and in time. 

\begin{table}[tb]

\caption{Sample results of the original 
algorithmic cooling \cite{AC-PNAS}
of 20 computation bits (and with $\ell =5$).}
\label{Tab:AC1}
\begin{center}

\begin{tabular}{|c|c|c|c|c|}
\hline 
$\epsilon_{initial}$&
$\epsilon_{desired}$&
$j_{f}$&
$N_{j_{f}}$&
$T_{j_{f}}$\\
\hline
\hline 
\multicolumn{1}{|c|}{}&
$8\epsilon_0$&
3&
140&
25$\times 10^{4}$\\
\cline{2-2} \cline{3-3} \cline{4-4} \cline{5-5} 
\multicolumn{1}{|c|}{$\epsilon_0 \ll 1$}&
$16\epsilon_0$&
4&
180&
125$\times 10^{4}$\\
%\cline{2-2} \cline{3-3} \cline{4-4} \cline{5-5} 
%\multicolumn{1}{|c|}{}&
%0.99&
%5&
%220&
%625$\times 10^{4}$\\
\hline 
%&
%0.5&
%6&
%260&
%3125$\times 10^{4}$\\
%\cline{2-2} \cline{3-3} \cline{4-4} \cline{5-5} 
%\multicolumn{1}{|c|}{0.01}&
%0.8&
%7&
%300&
%15625$\times 10^{4}$\\
%%\cline{2-2} \cline{3-3} \cline{4-4} \cline{5-5} 
%%\multicolumn{1}{|c|}{}&
%%0.9&
%%8&
%%340&
%%78125$\times 10^{4}$\\
%\hline
\end{tabular}
\end{center}
\end{table}

\begin{table}[tb]
\caption{Sample results of our 
improved algorithmic cooling of 20 computation bits.}
\label{Tab:AC2}
\begin{center}
\begin{tabular}{|c|c|c|c|c|}
\hline 
$\epsilon_{initial}$&
$\epsilon_{desired}$&
$J_{f}$&
$N_{J_{f}}$&
$T_{J_{f}}$\\
\hline
\hline 
\multicolumn{1}{|c|}{}&
$7.6 \epsilon_0$&
5&
30&
4040\\
\cline{2-2} \cline{3-3} \cline{4-4} \cline{5-5} 
\multicolumn{1}{|c|}{$\epsilon_0 \ll 1$}&
$  17.1 \epsilon_0$&
7&
34&
36440\\
%\cline{2-2} \cline{3-3} \cline{4-4} \cline{5-5} 
%\multicolumn{1}{|c|}{}&
%0.99&
%8&
%36&
%109340\\
\hline 
%&
%0.5&
%10&
%40&
%984140\\
%\cline{2-2} \cline{3-3} \cline{4-4} \cline{5-5} 
%\multicolumn{1}{|c|}{0.01}&
%0.8&
%12&
%44&
%8857340\\
%%\cline{2-2} \cline{3-3} \cline{4-4} \cline{5-5} 
%%\multicolumn{1}{|c|}{}&
%%0.9&
%%13&
%%46&
%%26572040\\
%\hline
\end{tabular}

\end{center}
\end{table}

\newpage

\section{Quantum bits and quantum gates in NMR quantum computing} 
\label{APP:qubits}

Although here we use the language of classical bits,
spins are actually quantum systems, 
and spin-half particles (two-level systems) 
are called quantum bits (qubits).
A molecule with
$n$ spin-half nuclei can represent an $n$-qubit 
computing device. 
The quantum computing device in this case is actually an ensemble
of many such molecules.
In ensemble NMR quantum computing~\cite{NMR-exp1,NMR-exp2}
each computer is represented by
a single molecule, such as the TCE molecule of Figure~3, 
and the qubits of the computer are represented by
the nuclear spins. 
A macroscopic number of identical molecules is available in a bulk
system, and these molecules act as though they are 
many computers performing the
same computation in parallel. 
The collection of many such molecules is put in a constant magnetic
field, so that a small majority of the 
spins (that represent a single qubit) are aligned with the direction
of that field. 
To perform a desired computation, the same sequence
of external pulses is applied to all the molecules/computers. 
Finally, a measurement
of the state of a single qubit 
is performed by summing over all computers/molecules
to read out the output on a particular qubit on all computers.
In practicable algorithmic cooling we know in advance the polarization
bias of the cooled bits after each step. 
In BMRVV algorithmic cooling we could also calculate the polarization
bias, but it becomes a very cumbersome process. Thus,
due to the complexity of the algorithm, the law of large
numbers is used, and calculation of the polarization bias is done only
after some particular steps called CUT in~\cite{AC-PNAS}.

For most purposes of this
current theoretical work, the gates we consider
are classical, and the spins can therefore
be considered as classical bits. Only in a lab,
the classical gates are implemented via the {\em available} 
quantum gates.
Then, the spins must be considered as qubits 
and the entire process is
a simple quantum computing algorithm.
However, unlike in other popular quantum algorithms, the use of 
quantum gates will not produce here any significant speed-up.

\section{Details of new bias calculations} \label{APP:epsilon-new}

as we have seen, the 3B-Comp   
can be built, for instance, via the following two gates:
\begin{enumerate}
\item
Use bit $C$ as a control, and bit $B$ as a target; apply a CNOT
(Controlled-NOT) operation: 
$C\rightarrow C$, $B\rightarrow B\oplus C$,
where $\oplus $ denotes a logical eXlusive OR (XOR).
\item
Use bit B as a control, 
and bits A and C as targets; apply a variant
of a C-SWAP operation: $A\rightarrow C\bar{B}+AB$, $B\rightarrow B$,
$C\rightarrow A\bar{B}+CB$; 
this means that $A$ and $C$ are swapped if $B=0$. 
\end{enumerate}
The effect of a 3B-Comp step in case it is applied
onto three bits with equal bias $\epsilon \ll 1$ is 
as follows: 
\begin{itemize}
\item
if $B=0$ after the CNOT operation,
the bias of $C$ at that stage is 
$(2\epsilon)/(1+\epsilon^{2})$ 
(this is the probability of $B=C=0$ given $B=C$,
all these calculated prior to the CNOT operation;
in other words, this is the 
probability of bit $C$ being 0, given that bit $B$ is
0 after the CNOT). Due to the CSWAP this bias is transfered 
to bit $A$;
\item
if $B=1$ after the CNOT operation,
the bias of $A$ is still $\epsilon$ after the CSWAP.
\end{itemize}
The new bias, $\epsilon_{\rm new}$, is the weighted
average 
$\epsilon_{\rm new}=
\left[ \frac{\left(1+\epsilon\right)^{2}}{4}
+ \frac{\left(1-\epsilon\right)^{2}}{4} \right]
\frac{2\epsilon}{1+\epsilon^{2}}+
\left[ \frac{1-\epsilon^{2}}{4} \times 2\right]
\epsilon$
of the two possible biases which is 
$\epsilon_{\rm new}=
\epsilon
\left[1+\frac{1-\epsilon^{2}}{2}\right] $,
which gives 
\begin{equation} 
\epsilon_{\rm new} \approx 3\epsilon/2 
\end{equation}
for small $\epsilon$. 

The following single operation can replace the CNOT+CSWAP, performing
a 3B-Comp, to cool bit $A$ (the heated 
bits $BC$ are not in the same state as for the previous 
3B-Comp, though):
\begin{eqnarray}
{\rm input:ABC} \ &  \quad \quad {\rm output:ABC} \nonumber \\
\quad \quad 000 \ & \rightarrow \quad \quad \quad 000   \nonumber \\
\quad \quad 001 \ & \rightarrow \quad \quad \quad 001   \nonumber \\
\quad \quad 010 \ & \rightarrow \quad \quad \quad 010   \nonumber \\
\quad \quad 011 \ & \rightarrow \quad \quad \quad 100   \nonumber \\
\quad \quad 100 \ & \rightarrow \quad \quad \quad 011   \nonumber \\
\quad \quad 101 \ & \rightarrow \quad \quad \quad 101   \nonumber \\
\quad \quad 110 \ & \rightarrow \quad \quad \quad 110   \nonumber \\
\quad \quad 111 \ & \rightarrow \quad \quad \quad 111   \nonumber
\end{eqnarray}

Note that reset bits can be used for the computation as
well. Thus the simplest algorithmic cooling 
can be obtained via 2 computing bits and
one reset bit. 

Note also that the strong tools of data compression can easily be used
to significantly improve the algorithms, 
for the price of dealing with more
complicated gates.

\section{A more space-efficient practicable algorithmic cooling} 
\label{APP:PAC2}

We consider here the case in which 
steps 1 and 2 
(the 3B-Comp and PT steps) in the outline 
of algorithmic cooling are combined into one generalized step of 
RPC. Then, the logical gates are
applied onto {\em all bits} in the system, 
that is computation and reset bits, to
push the entropy into the reset bits.
For comparison with the RPC and PAC1 (see
Figure~1)
we consider the cooling of a single bit.
% Step 3 means
% waiting without doing anything, for a time much
% longer than the relaxation
% time of the reset bits and much shorter than the relaxation time of
% the computation bits. 

Once we use the reset bits for the compression steps as well, 
replacing 
the 3B-Comp+PT steps by a generalized RPC, we can much 
improve the space complexity of the algorithm relative to PAC1.
We call this improved algorithm PAC2.

Whenever we used $2n$ bits in PAC1, 
namely $n$ computation bits, 
each one having a reset bit as a neighbour, 
we can now use exactly $n$ bits, 
namely $n-1$ computation bits plus one reset bit.
Let us explicitly show how this is done.
Let $\epsilon_0$ be the polarization bias of the reset bit.
In order to cool a single bit, say bit $A$, 
to $\epsilon_1$ start with two computation 
bits, $AB$ and one reset bit, $C$, 
perform PT($C\rightarrow B$) followed by
PT($B\rightarrow A$ to initiate bit $A$, 
RESET($C$) (by waiting), then perform
PT($C\rightarrow B$) to initiate bit $B$, and another RESET($C$). 
If the thermalization time of the compuation bits is sufficiently
large, we now have three bits with polarization bias $\epsilon_0$.
Now perform 3B-Comp to increase the
polarization of bit $A$. 

In order to cool one bit (say, bit $A$) to the second purification
level (polariation bias $\epsilon_{2}$), 
start with 4 computation bits $(ABCD)$, and one reset bit $E$. 
Perform PT steps sequentially (with RESET($E$) when needed) 
to initiate bits $ABC$, 
followed by a 3B-Comp on bits $ABC$ that
cools bit $A$ to a bias $\epsilon_{1}$. 
Then, perform PT steps sequentially (with RESET($E$) when needed) 
to initiate bits $BCD$, 
followed by a 3B-Comp on bits $BCD$ that
cools bit $B$ to a bias $\epsilon_{1}$. 
Next, perform PT steps sequentially (with RESET($E$) when needed) 
to initiate bits $CD$, followed by RESET($E$), and then  
followed by a 3B-Comp on bits $CDE$ that
cools bit $C$ to a bias $\epsilon_{1}$. 
Finally, apply 3B-Comp to bits $ABC$ to purify bit
$A$ to the second purification level 
$\epsilon_2 \approx (3/2)^2 \epsilon_0$. 
Clearly, the same strategy can be used to cool a single bit to 
$\epsilon_{j_F} \approx (3/2)^{j_F} \epsilon_0$ using
$2 J_F$ computation bits and a single reset bit. 
The total space complexity is therefore
\begin{equation} \label{EQ:PAC2-Algo}
2 J_F + 1
\end{equation}
Note that $(3/2)^4 = 81/16$ which is slightly larger than 5,
leading immediately to the 
numbers presented in Figure~1.
Cooling by a factor of 25 requires 17 spins here, 
while it requires 625 spins in RPC, 
or it requires repeating 
the experiment 625 times if no cooling is used.
 
We should mention, though, that the timing considerations for PAC2
are more demanding than the timing consideartions for PAC2 
due to the need of many more SWAP gates.  All the steps 
of the algorithm must be done within the relaxation time of the 
computation bits, and even more demanding, within the 
dephasing time (known as $T_2$) of the computation bits.

%%% END SUPPLEMENTARY MATERIAL


\begin{thebibliography}{10}

\bibitem{PT}Morris,~G.A.\ \& Freedman, R.\ Enhancement of nuclear magnetic
resonance signals by polarization transfer. \emph{J.\ Am.\ Chem.\ Soc.}\ \textbf{101,}
760--762 (1979). 

\bibitem{Sorensen}S\o rensen,~O.W.\ Polarization transfer experiments in high-resolution
NMR spectroscopy. \emph{Prog.\ Nuc.}\ \emph{Mag.}\ \emph{Res.\ spect.}\ \textbf{21,}
503--569 (1989). 

\bibitem{SV99}Schulman,~L.J.\ \& Vazirani,~U.\ Molecular scale heat engines
and scalable quantum computation. \emph{Proc.}\ \emph{31'st ACM STOC
(Symp.\ Theory of Computing)} 322--329 (1999). 

\bibitem{Griffin}Farrar,~C.T., Hall,~D.A., Gerfen,~G.J., Inati,~S.J.\ \& Griffin,~R.G.\ Mechanism
of dynamic nuclear polarization in high magnetic fields. \emph{J.\ Chem.\ Phys.}\ \textbf{114,}
4922-4933 (2001). 

\bibitem{AC-PNAS}Boykin,~P.O.\emph{,} Mor,~T.\emph{,} Roychowdhury,~V., Vatan,~F.\emph{\ }\&
Vrijen,~R.\ Algorithmic cooling and scalable NMR quantum computers.
\emph{Proc.\ Natl.\ Acad.}\ \emph{Sci}.\emph{\ }\textbf{99}, 3388--3393
(2002). 

\bibitem{book-on-NMR}Slichter,~C.P.\ 
\emph{Principles of Magnetic Resonance} (Springer-Verlag,
1990).

\bibitem{Cool-by-therm}Brassard,~G., Fernandez,~J.M., Laflamme,~R., 
Mor,~T.\ \& Weinstein,~Y.\ Experimental
cooling of NMR spins beyond Shannon's bound. manuscript in preparation.

\bibitem{YW}Weinstein,~Y.\ \emph{Quantum Computation and Algorithmic Cooling
by Nuclear Magnetic Resonance} M.Sc.\ Thesis (Technion, Haifa, 2003). 

\bibitem{BCS-Exp}Chang,~D.E., Vandersypen,~L.M.K.\ \& Steffen,~M.\ NMR implementation
of a building block for scalable quantum computation. \emph{Chem.\ Phys.\ Lett.\ }\textbf{338,}
337-344 (2001). 

\bibitem{Shor}Shor,~P.\ Polynomial-time algorithms for prime factorization and
discrete logarithms on a quantum computer. \emph{SIAM J.\ on Comp.}\ \textbf{26,}
1484--1509 (1997). 

\bibitem{NMR-exp1}Cory,~D.G., Fahmy,~A.F.\ \& Havel,~T.F.\ Ensemble quantum computing
by NMR spectroscopy. \emph{Proc.\ Natl.\ Acad.\ Sci.}\ \textbf{94,}
1634--1639 (1997). 

\bibitem{NMR-exp2}Gershenfeld,~N.A.\ \& Chuang,~I.L.\ Bulk spin-resonance quantum
computation. \emph{Science} \textbf{275,} 350--356 (1997). 

\bibitem{factoring-15}Vandersypen,~L.M.K.\ \emph{et al.}\ Experimental realization of
Shor's quantum factoring algorithm using nuclear magnetic resonance.
\emph{Nature} \textbf{414,} 883-887 (2001). 

\bibitem{scaling-problem1}Warren,~W.S.\ The usefulness of NMR quantum computing. \emph{Science}
\textbf{277,} 1688--1689 (1997). 

\bibitem{scaling-problem2}DiVincenzo,~D.P.\ Real and realistic quantum computers. \emph{Nature}
\textbf{393,} 113--114 (1998). 

%%% SUPP BIBLIOGRAPHY

% \bibitem{AC-PNAS}Boykin,~P.O.\emph{,} Mor,~T.\emph{,} Roychowdhury,~V., Vatan,~F.\emph{\ }\&
% Vrijen,~R.\ Algorithmic cooling and scalable NMR quantum computers.
% \emph{Proc.\ Natl.\ Acad.}\ \emph{Sci}.\emph{\ }\textbf{99}, 3388--3393
% (2002). 

% \bibitem{NMR-exp1}Cory,~D.G., Fahmy,~A.F.\ \& Havel,~T.F.\ Ensemble quantum computing
% by NMR spectroscopy. \emph{Proc.\ Natl.\ Acad.\ Sci.}\ \textbf{94,}
% 1634--1639 (1997). 

% \bibitem{NMR-exp2}Gershenfeld,~N.A.\ \& Chuang,~I.L.\ Bulk spin-resonance quantum
% computation. \emph{Science} \textbf{275,} 350--356 (1997). 

% \bibitem{factoring-15}Vandersypen,~L.M.K.\ \emph{et al.}\ Experimental realization of
% Shor's quantum factoring algorithm using nuclear magnetic resonance.
% \emph{Nature} \textbf{414,} 883-887 (2001). 

% \bibitem{scaling-problem1}Warren,~W.S.\ The usefulness of NMR quantum computing. \emph{Science}
% \textbf{277,} 1688--1689 (1997). 

% \bibitem{scaling-problem2}DiVincenzo,~D.P.\ Real and realistic quantum computers. \emph{Nature}
% \textbf{393,} 113--114 (1998). 

% \bibitem{SV99}Schulman,~L.J.\ \& Vazirani,~U.\ Molecular scale heat engines
% and scalable quantum computation. \emph{Proc.}\ \emph{31'st ACM STOC
% (Symp.\ Theory of Computing)} 322--329 (1999). 

% \bibitem{book-on-NMR}Slichter,~C.P.\ \emph{Principles of Magnetic Resonance} (Springer-Verlag,
% 1990).

% \bibitem{PT}Morris,~G.A.\ \& Freedman, R.\ Enhancement of nuclear magnetic
% resonance signals by polarization transfer. \emph{J.\ Am.\ Chem.\ Soc.}\ \textbf{101,}
% 760--762 (1979). 

% \bibitem{Griffin}Farrar,~C.T., Hall,~D.A., Gerfen,~G.J., Inati,~S.J.\ \& Griffin,~R.G.\ Mechanism
% of dynamic nuclear polarization in high magnetic fields. \emph{J.\ Chem.\ Phys.}\ \textbf{114,}
% 4922-4933 (2001). 

% \bibitem{Sorensen}S\o rensen,~O.W.\ Polarization transfer experiments in high-resolution
% NMR spectroscopy. \emph{Prog.\ Nuc.}\ \emph{Mag.}\ \emph{Res.\ spect.}\ \textbf{21,}
% 503--569 (1989). 

% \bibitem{Shor}Shor,~P.\ Polynomial-time algorithms for prime factorization and
% discrete logarithms on a quantum computer. \emph{SIAM J.\ on Comp.}\ \textbf{26,}
% 1484--1509 (1997). 

% \bibitem{Cool-by-therm}Brassard,~G., Fernandez,~J.M., Laflamme,~R., Mor,~T.\ \& Weinstein,~Y.\ Experimental
% cooling of NMR spins beyond Shannon's bound. manuscript on preparation.


% \bibitem{YW}Weinstein,~Y.\ \emph{Quantum Computation and Algorithmic Cooling
% by Nuclear Magnetic Resonance} M.Sc.\ Thesis (Technion, Haifa, 2003). 

% \bibitem{BCS-Exp}Chang,~D.E., Vandersypen,~L.M.K.\ \& Steffen,~M.\ NMR implementation
% of a building block for scalable quantum computation. \emph{Chem.\ Phys.\ Lett.\ }\textbf{338,}
% 337-344 (2001). 

\end{thebibliography}
\end{document}